\documentclass[pdflatex,sn-nature]{sn-jnl}

\usepackage{graphicx}
\usepackage{amsmath,amssymb,amsfonts}
\usepackage{booktabs}
\usepackage{xcolor}
\usepackage{siunitx}                
\usepackage[section]{placeins}      

\begin{document}

\title[Electrically switchable vacancy state]{Electrically switchable vacancy state revealed by in-operando positron experiments}

\author*[1,4]{\fnm{Ric} \sur{Fulop}}\email{ricfulop@mit.edu}
\author[1,4]{\fnm{Laurence} \sur{Lyons IV}}
\author[1,4]{\fnm{Robert} \sur{Nick}}
\author[2]{\fnm{Marc H.} \sur{Weber}}
\author[3]{\fnm{Ming} \sur{Liu}}
\author[4]{\fnm{Haig} \sur{Atikian}}
\author[4]{\fnm{Uwe} \sur{Bauer}}
\author[4]{\fnm{Alexander C.} \sur{Barbati}}
\author[1]{\fnm{Neil} \sur{Gershenfeld}}

\affil*[1]{\orgdiv{Center for Bits and Atoms}, \orgname{Massachusetts Institute of Technology}, \orgaddress{\city{Cambridge}, \state{MA}, \postcode{02139}, \country{USA}}}
\affil[2]{\orgdiv{Institute of Materials Research, Positron Beamline}, \orgname{Washington State University}, \orgaddress{\city{Pullman}, \state{WA}, \postcode{99164}, \country{USA}}}
\affil[3]{\orgdiv{Nuclear Reactor Program, Department of Nuclear Engineering}, \orgname{North Carolina State University}, \orgaddress{\city{Raleigh}, \state{NC}, \postcode{27695}, \country{USA}}}
\affil[4]{\orgname{General Flash}, \orgaddress{\street{640 Memorial Dr.}, \city{Cambridge}, \state{MA}, \postcode{02139}, \country{USA}}}

\abstract{Whether the flash state in electrically driven solids involves non-equilibrium defect production or is accounted for by Joule heating alone has been debated since 2010. Using positron annihilation spectroscopy on copper, we observe a fully reversible, electrically switchable vacancy population: the DBS $S$-parameter rises above baseline whenever applied current exceeds a critical density and returns on current removal. Positron lifetime spectroscopy independently confirms open-volume defect formation and reveals a void-to-cluster relaxation hierarchy. The current-induced vacancy concentration exceeds the thermal-equilibrium value at $352\,^\circ$C by $>10^{6}\times$, is present only while current is applied, and vanishes within minutes. The nucleation rate scales steeply with the applied current, connecting the minute-scale kinetics resolved here to the sub-second flash events observed in ceramic sintering. These results demonstrate current-induced Frenkel-pair production in a metal and identify a defect-mediated, non-equilibrium contribution to the flash state.}

\maketitle

When a modest electric field is applied to a heated solid, a sharp transition can occur: the electrical conductivity rises by orders of magnitude in seconds, and powder compacts densify to near-full density at temperatures hundreds of degrees below those required by conventional sintering. This ``flash'' phenomenon, discovered in ceramics by Cologna, Rashkova, and Raj in 2010\cite{cologna2010} and since reported across oxides, carbides, and metals\cite{biesuz2019,raj2012,bamidele2024,k.jha2016beyond,das2024reactive,raj2016analysis}, is accompanied by electroluminescence and anomalous phase transformations\cite{k.jha2016beyond}.

The years since have produced a sustained debate over its microscopic origin: whether the applied current acts solely through ordinary Joule heating or through a distinct non-equilibrium coupling to the lattice. Raj and coworkers first proposed a non-thermal contribution in 2011, suggesting that electric fields above a threshold nucleate a ``defect avalanche'' that enhances diffusion by raising the pre-exponential diffusion factor rather than lowering the activation barrier\cite{raj2011fields,francis2011forging}. Raj's 2012 Joule-heating analysis\cite{raj2012} reinforced this, showing that macroscopic Joule heating alone could reproduce the observed densification only if local grain-boundary temperatures exceeded the bulk by several hundred degrees---implausibly. Narayan\cite{narayan2013} proposed a parallel mechanism in 2013 in which defect segregation and selective Joule heating along dislocations and grain boundaries lead to local grain-boundary melting, distinct from the bulk-defect avalanche. The defect avalanche branch was subsequently developed by Raj and collaborators into a Frenkel-pair generation mechanism producing charge-neutral vacancies, interstitials, and electron--hole pairs\cite{raj2016analysis,k.jha2016beyond,jo2024}, and placed on an atomistic footing by Jongmanns, Raj, and Wolf\cite{jongmanns2018,jongmanns2020}, whose molecular dynamics showed that proliferation of phonons near the Brillouin-zone edge above the Debye temperature can produce vacancy--interstitial pairs at concentrations orders of magnitude above thermal equilibrium.

A competing family of models has argued that the macroscopic features of flash can be accounted for entirely by thermal runaway, without invoking any non-equilibrium defect generation. Todd et al.\cite{todd2015} showed that if the negative temperature coefficient of resistivity of 3YSZ is fed into a classical electrothermal model, the voltage, current, and specimen-temperature characteristics of the flash event are reproduced without any additional physics. Zhang et al.\cite{zhang2017} reached a similar conclusion for ZnO, and subsequent studies have corroborated the thermal-runaway picture for several ceramic systems. The two pictures predict the same macroscopic signatures---a sharp conductivity rise, densification at low furnace temperatures, and electroluminescence\cite{k.jha2016beyond}---but very different microstructures, and no experiment had directly observed defect populations evolving under current. The strongest indirect evidence for the field-driven-defect picture came from ex-situ microscopy of frozen samples: Jo et al.\cite{jo2024} found epitaxially coherent suboxide colonies in flashed 8YSZ, and Das et al.\cite{das2025} reported loss of cohesion in metals at current densities of 150--200~A/mm$^2$ without evidence of melting. But post-mortem observations cannot establish whether the defects they reveal are the cause of the flash state or a frozen-in consequence of it.

\begin{figure}[!t]
\centering
\includegraphics[width=\textwidth]{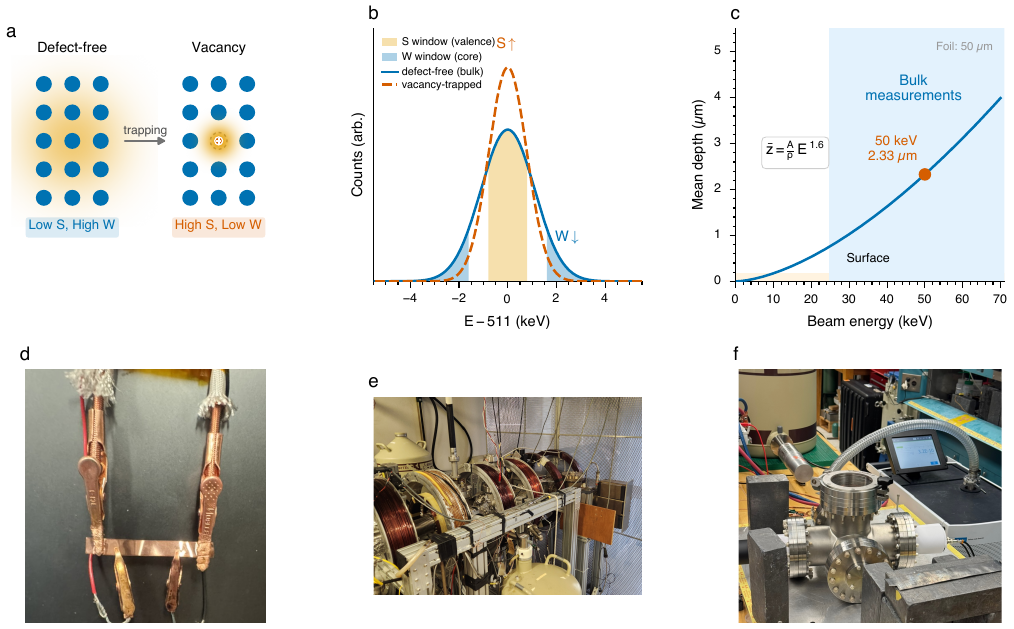}
\caption{\textbf{Positron DBS/PALS technique and experimental setup.} \textbf{a}, Positron trapping concept. In a defect-free copper lattice the thermalised positron (orange cloud) is delocalised over a Bloch state with a thermal de~Broglie wavelength of several nanometres (thousands of atomic sites; the schematic shows only a few unit cells) and annihilates predominantly with valence electrons, yielding a low-$S$, high-$W$ signature. At an open-volume defect (right; dashed circle marks the missing atom) the positron is localised in the vacancy and samples a different electron-momentum distribution, yielding a high-$S$, low-$W$ signature. \textbf{b}, The $S$ and $W$ windows on the 511~keV annihilation photopeak. The electron--positron pair momentum Doppler-broadens the line; the fraction of counts in the central low-momentum window is $S$ (orange shading, valence electrons) and the fraction in the high-momentum wings is $W$ (blue shading, core electrons). Solid blue curve: defect-free bulk; dashed orange curve: positron trapped in a vacancy (both curves normalised to equal total area). Trapping narrows the line because the localised positron sees fewer core electrons, driving $S\uparrow$ and $W\downarrow$. \textbf{c}, Mean positron implantation depth $\bar{z}$ in copper as a function of beam energy, $\bar{z} = (A/\rho)E^{1.6}$ with $A = 4.0$~\si{\micro\gram/\centi\metre\squared}/keV$^{1.6}$ and $\rho = 8.96$~g/cm$^3$. At 50~keV, $\bar{z} = 2.33$~\si{\micro\metre} (orange dot), well within the 50~\si{\micro\metre} foil (shaded bulk region); the WSU DBS experiments operate in this bulk-sensitive regime. \textbf{d}, Four-point-probe sample holder: a 50~\si{\micro\metre}-thick electrolytic Cu foil is clamped between spring-loaded alligator-clip terminations on 3~mm Cu bus bars; this fixture is the same at both facilities. \textbf{e}, WSU 70~keV variable-energy positron beamline, showing the solenoid transport magnets that deliver the implanted beam to the sample at the end of the line (HpGe detector not in view). \textbf{f}, Assembled NC State bulk positron system used for in-situ measurement: a vacuum chamber with two re-entrant flanges and power/signal feedthroughs; two Hamamatsu PMTs for PALS and an ORTEC HPGe detector for DBS; the copper foil specimen sits ${\sim}2$~mm from a $^{22}$Na source deposited on a single-crystal silicon wafer.}
\label{fig:technique}
\end{figure}

Resolving this requires a non-destructive, in-situ probe that operates under current and temperature and is sensitive to vacancy-type defects at parts-per-million levels. Positron Doppler broadening spectroscopy (DBS) and positron annihilation lifetime spectroscopy (PALS) meet all three criteria\cite{krause1999,tuomisto2013,hautojarvi1995}. In a defect-free metal the thermalised positron is delocalised over a Bloch state with a thermal de~Broglie wavelength of several nanometres (thousands of atomic sites); at an open-volume defect it is trapped and samples a different electron-momentum distribution (Fig.~\ref{fig:technique}a,b). In DBS, the lineshape parameter $S$ quantifies the low-momentum fraction of positron--electron annihilation events and rises at defects where fewer core electrons are available; the complementary parameter $W$ measures the high-momentum fraction and drops. The slope of the $S$--$W$ correlation identifies the defect type, and the two-state trapping model relates $\Delta S$ to the vacancy concentration. PALS independently measures defect size via the positron lifetime, which increases with the open volume of the trapping site.

We chose copper as the test material because it provides the most stringent null hypothesis for a thermal explanation. The thermal-vacancy sigmoidal onset in copper occurs near ${\sim}550$~$^\circ$C\cite{fluss1980,hehenkamp1986} ($H_f = 1.31 \pm 0.05$~eV\cite{fluss1980,simmons1963,hehenkamp1986}), and the Stage III defect-recovery spectrum has been exhaustively characterized by positron annihilation\cite{mantl1978,ehrhart1991}. Any sustained defect signal at temperatures well below ${\sim}550$~$^\circ$C must therefore arise from the applied current rather than from equilibrium thermodynamics.

A central challenge for in-operando measurement is that flash is ordinarily abrupt: at the current densities used in sintering, the transition completes in tens to hundreds of milliseconds. However, because defect nucleation rates above threshold are expected to scale steeply with applied current\cite{jongmanns2018}, the kinetics can be placed in any temporal window by tuning the overdrive ratio $J/J_c$ (the ratio of the applied current density $J$ to the critical current density threshold $J_c$). Here we exploit this by operating at $J/J_c \approx 1.12$--$1.15$, only slightly above threshold, which stretches the normally sub-second process to minutes, well within the temporal resolution of a positron beam. Experiments were conducted at two U.S.\ positron facilities using a common four-point-probe sample geometry (Fig.~\ref{fig:technique}d): time-resolved DBS at the Washington State University 70~keV line-beam facility (17.5~hr across five campaigns, Fig.~\ref{fig:technique}e) and PALS at the NC State PULSTAR positron facility (${\sim}20$~hr, Fig.~\ref{fig:technique}f). Together these experiments show that the applied current reversibly switches the vacancy population of bulk copper as it enters flash, and that the steep rate scaling connects our minute-scale kinetics to the sub-second flash observed at higher overdrive ratios.

\section*{Results}

\subsection*{Defect-free baseline}

To establish the defect-free reference, variable-energy DBS scans (0--70~keV) were performed on a Cu(100) single crystal etched in nitric acid (132 points, 25.4~hr), yielding a bulk ($E > 25$~keV) reference $S_\text{ref} = 0.3932 \pm 0.002$ and $W_\text{ref} = 0.1343 \pm 0.001$ (Fig.~\ref{fig:threshold}b; the trapping concept is illustrated in Fig.~\ref{fig:technique}a). All subsequent $S$ values are compared to this single-crystal reference: $S/S_\text{ref} = 1$ indicates a defect-free bulk; values above unity indicate open-volume defects. A depth scan on an as-received copper foil (an electrolytic Cu foil prior to any thermal or electrical processing; 56 points spanning 0--70~keV, run CuTest6/7) gives a bulk $S = 0.398$ ($S/S_\text{ref} = 1.012$), reflecting residual cold-work defects that serve as the starting condition for the current-cycling experiments. All experimental runs and their lab identifiers (``CuTest4'', etc., referenced below) are catalogued in Extended Data Table~\ref{tab:runs}.

\subsection*{Reversible S-parameter switching under current cycling}

\begin{figure}[tbp]
\centering
\includegraphics[width=\textwidth]{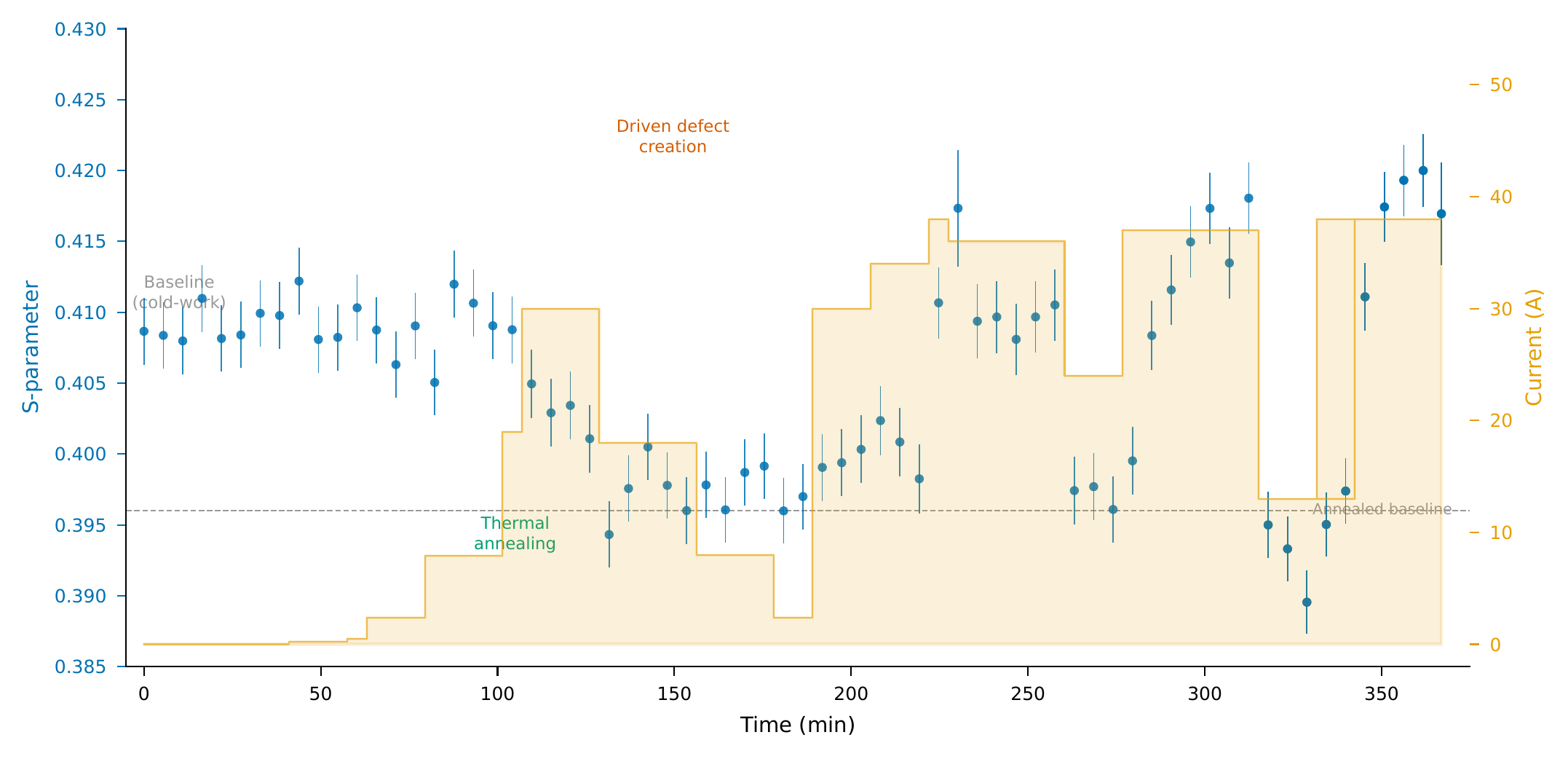}
\caption{\textbf{Reversible vacancy switching under current cycling.} S-parameter (blue, left axis) and applied current (amber bars, right axis) versus time for the CuTest4 experiment (67 DBS measurements, 50~keV, WSU). Three regimes are evident: initial cold-work baseline ($S \sim 0.409$), thermal annealing under moderate current ($S \to 0.396$), and reversible current-induced defect creation above ${\sim}30$~A ($S \to 0.42$) with recovery below ${\sim}20$~A, repeated four times. Dashed line: annealed baseline. Error bars: $\pm1\sigma$ from counting statistics.}
\label{fig:timeseries}
\end{figure}

Figure~\ref{fig:timeseries} shows the DBS $S$-parameter and applied current versus time for the full CuTest4 experiment (67 measurements at 50~keV with a mean interval between measurements, or cadence, of ${\sim}5.5$~min), together forming the central observation of this work. Three successive regimes are resolved: an initial cold-work baseline at 0~A; a thermal-annealing regime as the current ramps through 0.2--30~A; and a reversible switching regime at currents above ${\sim}30$~A.

During the initial 0~A interval (0--40~min), the as-received foil sits at $S = 0.409 \pm 0.001$ ($S/S_\text{ref} = 1.040$, $N = 8$), consistent with residual cold-work defects. Over the subsequent thermal-annealing interval (${\sim}40$--180~min, $I$ ramped through 0.2--30~A), $S$ falls to ${\sim}0.396$ ($S/S_\text{ref} = 1.007$), as predicted by the Stage~III thermal recovery of Mantl and Triftsh\"auser\cite{mantl1978}; this establishes the annealed reference within 0.7\% of single crystal, $S_\text{annealed} = 0.396 \pm 0.003$ ($N = 9$, all points with $I \leq 13$~A after $t = 100$~min).

At currents above this annealing range, the $S$-parameter switches reversibly between the annealed value and an elevated value, tracking the applied current. Each time the current exceeds the critical current density $J_c$ for defect switching in this foil geometry (identified here as $J_c \approx 110$~A/mm$^2$, corresponding to $I_c \approx 33$~A for the $0.30~\mathrm{mm}^2$ cross-section, and refined quantitatively in the next subsection), $S$ rises to 0.41--0.42 ($S/S_\text{ref} = 1.04$--$1.07$, a 4--7\% enhancement over single crystal); each time the current drops below ${\sim}20$~A, $S$ returns to ${\sim}0.396$ ($S/S_\text{ref} = 1.007$). Four independent switching cycles are captured in CuTest4: at 30~A ($t \approx 192$~min; marginal, near threshold), 34--38~A ($t \approx 209$~min), 37~A ($t \approx 280$~min), and 38~A ($t \approx 335$~min). Of these, the 37~A and 38~A ramps are fully resolved nucleation transients: $S$ rises from 0.400 to 0.418 over ${\sim}30$~min at 37~A and from 0.395 to 0.420 over ${\sim}30$~min at 38~A, separated by a deliberate reduction to 13~A that allowed the vacancy population to recombine back to the annealed baseline before the next ramp was initiated. Because these two transients are reproduced in the same specimen during the same run, the experiment is internally controlled: the same thermal history that produces Stage~III recovery at moderate current produces the elevated $S$ at high current, so the high-current increase cannot be attributed to residual cold work or to a one-off instrumental shift.

The magnitude of this switching signal ($\Delta S = 0.024$) is substantially larger than any plausible noise floor on the measurement. A dedicated baseline-stability run (CuTest3: 46 consecutive DBS measurements at 50~keV with no current) yields $S = 0.409 \pm 0.005$, with the scatter dominated by counting statistics (per-point $\delta S \sim 0.007$). The CuTest4 per-point uncertainties are smaller ($\delta S \sim 0.0024$) because of the longer integration time, so the switching amplitude exceeds the per-point counting uncertainty by approximately an order of magnitude and is several times the long-term scatter of the no-current baseline.

\subsection*{Independent threshold confirmation from resistance}

We adopt as our operational definition of $J_c$ the fit-derived threshold from the nucleation-rate analysis (next subsection): $J_c \approx 110$~A/mm$^2$ ($I_c = 33 \pm 3$~A for the $0.30~\mathrm{mm}^2$ cross-section; the uncertainty spans the marginal $S$-rise at 30~A and the fully sustained switching at 36~A). This corresponds to $E_c \approx 2.7$~V/m (${\sim}0.027$~V/cm) and a threshold sample voltage $V_c \approx 80$~mV across the 30~mm gauge. For context, ceramic flash sintering typically requires 50--1000~V/cm; in metals the field is orders of magnitude lower but the current density is correspondingly higher because of the low resistivity.

\begin{figure}[tbp]
\centering
\includegraphics[width=\textwidth]{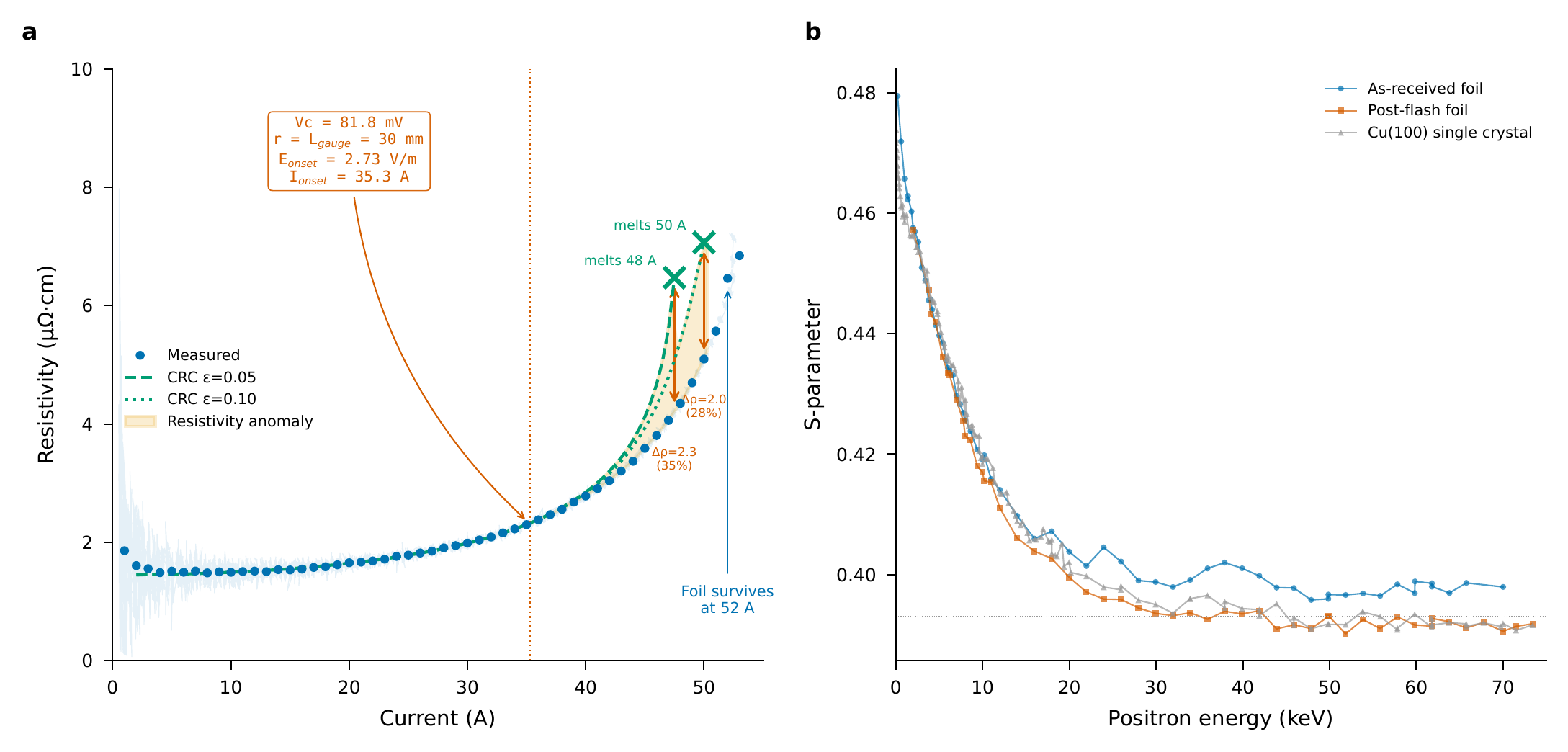}
\caption{\textbf{Threshold corroboration and depth profiles.} \textbf{a}, Four-point resistivity versus current during a 0--53~A ramp. Measured resistivity (blue) is compared with two CRC thermal predictions bracketing the plausible surface emissivity of the foil ($\varepsilon = 0.05$ and $0.10$). The thermal model predicts melting at 48--50~A; the foil remains intact up to 52~A. \textbf{b}, $S$-parameter versus positron beam energy for three specimens: the as-received electrolytic Cu foil (blue), the same foil after the complete CuTest4/5 current-cycling run (red; referred to in the text as the post-flash foil), and the Cu(100) single-crystal reference (grey). In the bulk ($E > 25$~keV), the post-flash foil and the single crystal agree within ${<}1\%$, so the current-induced vacancy population is transient and leaves no detectable residual defect signal on the scale resolved by DBS.}
\label{fig:threshold}
\end{figure}

Four-point resistivity measured simultaneously during an initial current ramp (0--53~A over 140~s, Fig.~\ref{fig:threshold}a) rises monotonically with current, increasing slowly below ${\sim}20$~A and with steeper, concave-upward curvature above ${\sim}30$~A. A calibrated 1D thermal model\cite{matula1979,crc97} that balances Joule heating against radiation and axial conduction to the clips (surface emissivity $\varepsilon = 0.05$--0.10, the range appropriate for polished-to-lightly-oxidized copper) predicts the CRC thermal resistivity of the foil should exceed the measured values by 28--35\% at the melt point, and predicts melting at 48--50~A; in the experiment the foil remains intact (no melting observed) up to 52~A. The onset of this deviation from the thermal model coincides with the DBS switching threshold, so two independent observables---the positron $S$-parameter and the four-point resistivity---point to the same critical current.

\subsection*{Post-flash recovery}

After the full CuTest4/5 current-cycling run (67~measurements covering 0--38~A in CuTest4 followed by the 38~A~$\to$~0~A decay of CuTest5), a 49-point depth scan of the same foil---referred to from here on as the post-flash foil---converges to a bulk $S = 0.392$ ($S/S_\text{ref} = 0.997$), indistinguishable from the Cu(100) single-crystal reference (Fig.~\ref{fig:threshold}b). The current-induced vacancy population is therefore transient: once the current is removed the foil recovers to the single-crystal reference within the DBS uncertainty.

\subsection*{Nucleation rate scaling with applied current}

\begin{figure}[tbp]
\centering
\includegraphics[width=\textwidth]{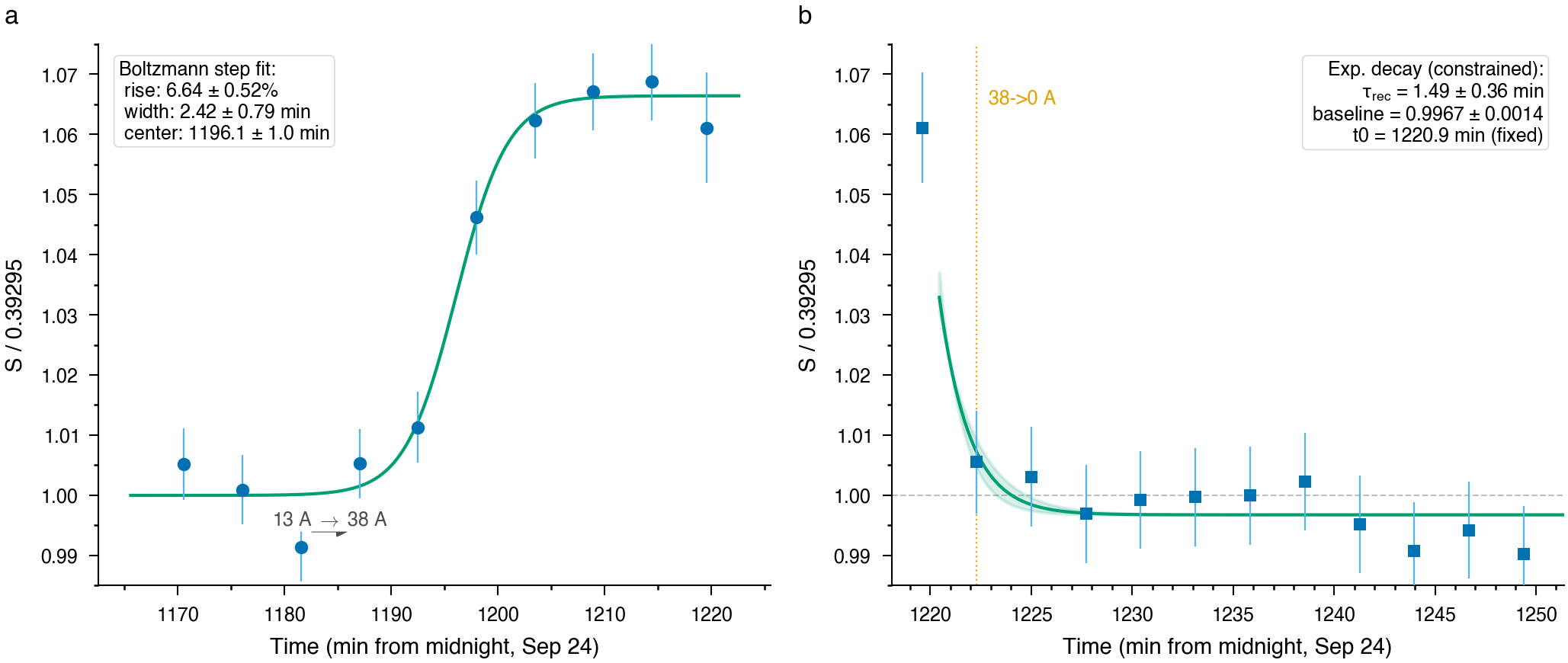}
\caption{\textbf{Nucleation and recombination kinetics.} \textbf{a}, Boltzmann step fit to the final 13$\to$38~A nucleation transient (CuTest4) of the form $S(t) = S_\text{low} + \Delta S/(1+\exp(-(t-t_0)/w))$. Rise $\Delta S/S_\text{low} = 6.64 \pm 0.52\%$; Boltzmann width $w = 2.42 \pm 0.79$~min. Note: $w$ is the Boltzmann-sigmoidal width and is distinct from the exponential nucleation time constant $\tau_\text{nuc}$ reported in the main text; the two are different functional-form parameters for the same rising transient. \textbf{b}, Constrained exponential decay fit to CuTest5 (38$\to$0~A) with recombination time constant $\tau_\text{rec} = 1.49 \pm 0.37$~min. The $1\sigma$ envelope (shaded) reflects the uncertainty on $\tau_\text{rec}$. Defect recovery is faster than the measurement cadence. Error bars: $\pm1\sigma$.}
\label{fig:kinetics}
\end{figure}

The two fully resolved nucleation transients yield exponential time constants $\tau_\text{nuc} \approx 12.1$~min at 37~A ($J/J_c = 1.12$) and $\tau_\text{nuc} \approx 6.7$~min at 38~A ($J/J_c = 1.15$), where $\tau_\text{nuc}$ is the time constant of an exponential approach to steady state (distinct from the Boltzmann width $w = 2.42 \pm 0.79$~min reported in the Fig.~\ref{fig:kinetics}a caption; see Methods for both fit forms and their relationship). A 3\% increase in current density nearly doubles the nucleation rate, consistent with a steep scaling rate~$\propto \exp(\alpha\,J/J_c)$ with $\alpha \approx 20$ ($\pm 30\%$ from two-point fitting; see Methods). Taking this scaling as illustrative rather than established, extrapolation to $J/J_c = 1.0$ gives $\tau_\text{nuc} \approx 130$~min, consistent with the marginal behaviour at 30--34~A; $J/J_c = 1.5$ gives $\tau_\text{nuc} \approx 400$~ms (factor-of-8 uncertainty from the $\alpha$ spread); and $J/J_c = 2$ gives sub-millisecond. If this scaling holds at higher overdrive, it would reconcile the minute-scale kinetics resolved here (deliberately only slightly above threshold) with the sub-second flash events reported in ceramic sintering at $J/J_c \approx 2$--5, where 25--35~mol\% densification occurs in ${\sim}100$~ms. The bridge to ceramic flash is a hypothesis consistent with the present data rather than an established result; we develop it in a companion preprint\cite{fulop2026}.

\subsection*{Rapid defect annihilation}

A dedicated decay measurement (CuTest5: 38~A $\to$ 0~A, cadence ${\sim}2.5$~min) shows $S$ dropping from 0.417 to ${\sim}0.391$ within the first measurement interval and remaining at the annealed baseline for the following 25~min (Fig.~\ref{fig:kinetics}b). Because the decay is faster than the cadence---$\tau_\text{rec} < 3$~min---the current-induced vacancy population is not permanent damage but a non-equilibrium population sustained only by the applied current.

\subsection*{PALS confirmation (NC State PULSTAR facility)}

To test the DBS findings independently and at a second facility, positron annihilation lifetime spectroscopy (PALS) was performed at the NC State PULSTAR positron facility on a separate set of electrolytic Cu foils of the same specification and in the same four-point-probe fixture geometry. Spectra were acquired at four applied-current conditions---an initial room-temperature no-current baseline, 38~A (the ``flash'' condition), 13~A (the partially annealed condition), and a post-flash room-temperature measurement after the 38~A and 13~A exposures---so that both the same current levels used in the DBS experiment and the recovery of the foil between exposures could be examined on the same specimen; the PALS setup is shown in Fig.~\ref{fig:technique}f and Extended Data Fig.~\ref{fig:pals_setup}, with the raw lifetime spectra reproduced in Extended Data Fig.~\ref{fig:pals_spectra}.

Table~\ref{tab:pals} summarizes the three-component fits to these spectra, obtained with LT10\cite{dryzek1996}---a standard positron-lifetime deconvolution program widely used in the positron-annihilation community. Because the copper foil was positioned ${\sim}2$~mm from a $^{22}$Na source mounted on single-crystal silicon, approximately half of the annihilation signal originates from the silicon substrate; since this silicon contribution is the same for every applied-current condition, the changes in the fit parameters between conditions are attributable to the copper foil. The fitted $\tau_1 \approx 213$~ps reflects an unresolved convolution of the bulk-copper lifetime (${\sim}110$~ps) and the bulk-silicon lifetime (${\sim}218$~ps): given the dominant silicon contribution and the ${\sim}300$~ps instrumental timing resolution, the 110~ps bulk-Cu component cannot be freely resolved as a separate lifetime in our fits.

\begin{table}[ht]
\centering
\caption{\textbf{PALS lifetime decomposition (NC State).} Three-component fits using LT10. $\tau_i$ are the fitted lifetimes and $I_i$ the corresponding intensities; ``Baseline (RT)'' and ``Post-flash (RT)'' denote room-temperature, no-current measurements before and after the 38~A/13~A exposures, respectively; $\chi^2_r$ is the reduced chi-squared of the fit. Changes between conditions are attributable to the copper foil because the silicon contribution to the signal is the same for every condition.}
\label{tab:pals}
\begin{tabular}{@{}lcccccccc@{}}
\toprule
\textbf{Condition} & $\tau_1$ (ps) & $I_1$ (\%) & $\tau_2$ (ps) & $I_2$ (\%) & $\tau_3$ (ns) & $I_3$ (\%) & Counts ($\times 10^6$) & $\chi^2_r$ \\
\midrule
Baseline (RT) & $213 \pm 1$ & 83.7 & $526 \pm 11$ & $14.5 \pm 1.1$ & $2.11 \pm 0.11$ & $1.78 \pm 0.23$ & 1.71 & 1.01 \\
38~A (flash)  & $212 \pm 1$ & 86.1 & $553 \pm 3$  & $11.8 \pm 0.2$ & $3.01 \pm 0.06$ & $2.08 \pm 0.10$ & 1.98 & 0.98 \\
13~A          & $206 \pm 2$ & 81.5 & $546 \pm 11$ & $16.7 \pm 1.3$ & $2.77 \pm 0.18$ & $1.79 \pm 0.28$ & 0.66 & 0.80 \\
Post-flash (RT) & $214 \pm 1$ & 82.2 & $552 \pm 9$ & $15.9 \pm 1.0$ & $2.66 \pm 0.11$ & $1.95 \pm 0.21$ & 1.49 & 1.02 \\
\bottomrule
\end{tabular}
\end{table}

\begin{figure}[t]
\centering
\includegraphics[width=\textwidth]{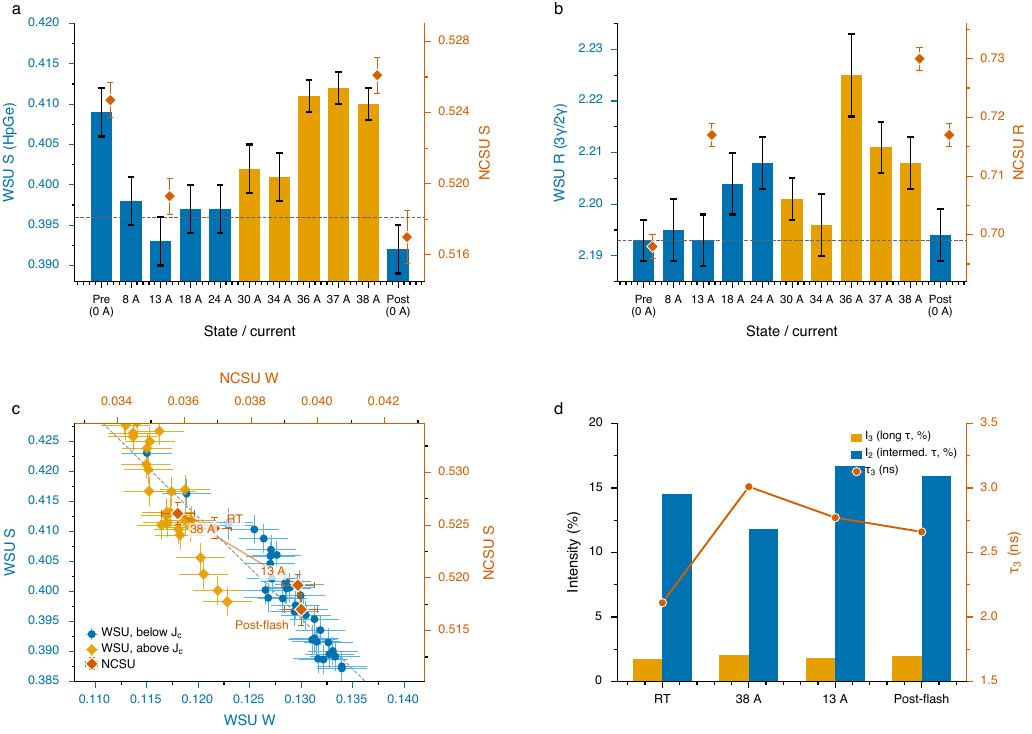}
\caption{\textbf{DBS and PALS confirmation.} \textbf{a}, WSU CuTest4 $S$-parameter versus physical state: ``Pre (0~A)'' is the initial cold-worked baseline, ``8~A''--``38~A'' are the mean $S$ values at each step of the CuTest4 ramp, and ``Post (0~A)'' is the post-flash foil after the full CuTest4+CuTest5 cycle (single-crystal quality). NC State DBS results for the four analogous states (RT, 13~A, 38~A, post-flash) are plotted as orange diamonds against a facility-matched right-hand $S$ axis. The left and right scales differ by a facility-specific additive constant because roughly half of the NC State annihilation signal comes from the silicon substrate carrying the $^{22}$Na source, which raises the apparent absolute $S$. Both facilities show the same pattern: $S$ rises above the annealed baseline at 38~A, sits at or below baseline at 13~A and after the full cycle (``Post''), and the WSU cold-worked ``Pre'' bar lines up with the NC State RT baseline. \textbf{b}, $R$-parameter ($3\gamma/2\gamma$ ratio) on the same dual-axis scheme, showing an elevated positronium signal at high current in both datasets, consistent with the formation of large open-volume features at 38~A. \textbf{c}, WSU $S$--$W$ correlation from CuTest4 (blue = below $J_c$, orange = above $J_c$) with the NC State $S$--$W$ data overlaid as vermillion diamonds. The top and right secondary axes show the NC State $W$ and $S$ scales respectively; the two axis systems are calibrated by a linear (affine) transformation anchored on the flash and annealed states, so each NC State diamond visually lies on top of the corresponding WSU cluster. The NC State RT baseline sits at the WSU cold-worked position (between the annealed and flash clusters); 38~A is in the above-$J_c$ flash cluster; 13~A is intermediate; and the post-flash point falls in the below-$J_c$ annealed cluster, confirming the recovery. \textbf{d}, NC State PALS lifetime decomposition: $I_3$ (long-lifetime intensity, orange bars) peaks at 38~A, consistent with enhanced positronium formation at the internal surfaces of large open-volume features; $I_2$ (intermediate-lifetime intensity, blue bars) rises at 13~A and post-flash, consistent with a defect relaxation hierarchy. $\tau_3$ (line) tracks the long-lifetime component.}
\label{fig:pals}
\end{figure}

At 38~A, the long-lifetime component increases ($\tau_3$: $2.11 \to 3.01$~ns; $I_3$: $1.78 \to 2.08$\%), indicating the formation of large open-volume features. The simultaneous increase of both $\tau_3$ and $I_3$---parameters that are typically anti-correlated in lifetime fitting---indicates a genuine change in the defect population rather than a fitting artefact; the product $\tau_3 \times I_3$ rises monotonically from 3.75 (baseline) to 6.26 (38~A), consistent with the DBS measurement. Figure~\ref{fig:pals}a shows the WSU CuTest4 $S$-parameter at each applied-current level (50~keV, left axis, blue bars) together with the four independent NC State measurement states shown as diamonds on a facility-matched right-hand axis. The WSU bar chart is now organised by physical state rather than current alone: the leftmost ``Pre (0~A)'' bar is the initial cold-worked baseline before any thermal anneal, the 8--38~A bars correspond to the CuTest4 current ramp, and the rightmost ``Post (0~A)'' bar is the post-flash foil after the full CuTest4+CuTest5 cycle (single-crystal quality, $S = 0.392$). Each NC State diamond sits next to the WSU bar representing the matching physical state: RT $\leftrightarrow$ Pre; 13~A $\leftrightarrow$ 13~A; 38~A $\leftrightarrow$ 38~A; post-flash $\leftrightarrow$ Post. The absolute NC State $S$ values are offset from the WSU values by a facility-specific additive constant---approximately half of the NC State annihilation signal originates from the silicon substrate on which the $^{22}$Na source is deposited, which raises the apparent absolute $S$---so the left and right axes use matched but distinct scales. Under this normalisation, both facilities show the same pattern: $S$ rises above the annealed baseline at 38~A (flash), sits at or below the baseline at 13~A (partial anneal) and after the full cycle (single-crystal-quality ``Post''), and the WSU ``Pre'' cold-worked state lines up with the NC State RT baseline. Fig.~\ref{fig:pals}b shows the corresponding $R$-parameter ($3\gamma/2\gamma$ ratio) on the same dual-axis scheme; the NC State $R$-parameter rises from 0.698 (RT) to 0.730 at 38~A and falls to 0.717 after the current is removed, independently confirming an enhanced open-volume contribution at high current, with the WSU $R$ bars tracking the same pattern. When the current is reduced from 38~A to 13~A, $I_3$ decreases while $I_2$ increases from 11.8\% to 16.7\%: spectral weight transfers from the long-lifetime component (associated with positronium formation at internal surfaces created by large open-volume features) to the intermediate-lifetime component, revealing a relaxation hierarchy in which the largest defects collapse first (Fig.~\ref{fig:pals}d). We note that $I_2$ may include a small contribution from the geometry of the $^{22}$Na source. The corresponding $S$--$W$ plot (Fig.~\ref{fig:pals}c) shows the CuTest4 points clustering distinctly at each current condition.

\subsection*{S--W correlation}

The S--W trajectory from CuTest4 (Fig.~\ref{fig:pals}c) shows reversible migration between two clusters: high-current points at $(S, W) \approx (0.42, 0.115)$, and low-current points at $(0.396, 0.131)$. The slope is characteristic of vacancy-type open-volume defects\cite{hautojarvi1995} and distinct from the single-crystal and as-received references, confirming vacancy character.

\subsection*{Trap concentration}
In the two-state trapping model\cite{krause1999,mantl1978}, the measured $S$-parameter is the weighted average of the bulk and defect-trapped contributions:
\begin{equation}
S = (1 - \eta)\,S_b + \eta\,S_d, \qquad \eta = \frac{\mu\,C_v}{\mu\,C_v + \lambda_b},
\label{eq:trapping}
\end{equation}
where $\eta$ is the trapping fraction, $\mu$ is the specific positron trapping rate, $C_v$ the vacancy concentration, $\lambda_b = 1/\tau_b$ the bulk annihilation rate, and $S_b$ and $S_d$ the lineshape parameters for the bulk and defect-trapped states, respectively. Rearranging gives
\begin{equation}
C_v = \frac{\lambda_b}{\mu}\,\frac{S - S_b}{S_d - S}.
\label{eq:concentration}
\end{equation}
If current-induced vacancies are created, $S$ should rise above the defect-free baseline $S_b$ towards $S_d$; if the flash state is purely thermal in origin, $S$ should instead remain at $S_b$ for foil temperatures below ${\sim}550\,^\circ$C.

With $\mu = 1.1 \times 10^{14}$~s$^{-1}$, $\tau_b = 110$~ps, and DBS contrast $(S_d - S_b)/S_b = 0.07$--$0.10$, Eq.~\ref{eq:concentration} gives an upper-bound $C_v \lesssim 130$--$530$~ppm for $\Delta S = 0.024$ (upper-bound because the true $S_d$ may exceed the reported cluster/void contrasts; see Methods). The data force $(S_d - S_b)/S_b > 0.06$, above the ${\approx}0.04$ value typical for monovacancies in copper\cite{hautojarvi1995,mantl1978,dlubek1979}, indicating a population dominated by vacancy clusters and voids, consistent with the PALS lifetime hierarchy. The foil temperature at 38~A in a matching fixture was $352 \pm 2\,^\circ$C by spot-welded Type~K thermocouple, $358 \pm 5\,^\circ$C by calibrated Optris LWIR, and $382 \pm 15\,^\circ$C by two-colour pyrometer (see Methods)---$352\,^\circ$C (625~K), well below the ${\sim}550\,^\circ$C thermal-vacancy onset. At 625~K, $C_\text{eq} \approx 3$--$5 \times 10^{-5}$~ppm ($H_f = 1.28$~eV\cite{simmons1963}; $H_f = 1.31$~eV\cite{fluss1980,hehenkamp1986}), so the current-induced population exceeds the thermal-equilibrium value by ${\sim}10^{6}\times$---a margin near the upper end of positron-DBS sensitivity, but one that cannot be accounted for by Joule heating.

\FloatBarrier
\section*{Discussion}

Our measurements identify a reversible, current-induced vacancy population in bulk copper that switches on and off with the applied current at a temperature well below the thermal-vacancy onset. Four independent lines of evidence converge: the DBS $S$-parameter switches reversibly through four cycles; the four-point resistivity deviates from the thermal baseline at the same critical current; PALS at a second facility confirms open-volume defect formation and a cluster-to-void hierarchy; and depth profiles after current removal recover to the single-crystal reference. A follow-up experiment measured the foil temperature at 38~A as $352\,^\circ$C (thermocouple), $358\,^\circ$C (LWIR), and $382\,^\circ$C (pyrometer)---${\sim}198\,^\circ$C below the thermal-vacancy onset.

These results allow us to distinguish between the two mechanistic pictures that have dominated the flash literature. The thermal-runaway family\cite{todd2015,zhang2017} predicts negligible excess vacancies below ${\sim}550\,^\circ$C; we instead observe a defect population $>\!10^{6}\times$ the equilibrium value at $352\,^\circ$C. The field-driven-defect family\cite{raj2011fields,francis2011forging,raj2012,narayan2013,raj2016analysis,jongmanns2018,jongmanns2020} predicts vacancy creation above the Debye temperature with reversible, threshold-dependent kinetics; the data reproduce each of these features. The kinetic signatures we resolve---sigmoidal nucleation, sub-minute recombination, a current-density threshold, and a cluster-to-void hierarchy---provide a template for interpreting the frozen microstructures reported previously: the suboxide colonies in flashed 8YSZ\cite{jo2024}, the loss of cohesion in metals below the melting point\cite{das2025}, the electroluminescence\cite{k.jha2016beyond}, and the current-density onsets\cite{bamidele2024} all map naturally onto a non-equilibrium vacancy population. A phonon-mediated critical-voltage interpretation of $V_c \approx 80$~mV and the overdrive-ratio scaling framework are developed in a companion preprint\cite{fulop2026}.

A thermal origin for the observed vacancy population can be excluded quantitatively. Fluss et al.\cite{fluss1980} measured the Doppler-broadening lineshape parameter of copper from 25 to 1040~$^\circ$C, and Hehenkamp et al.\cite{hehenkamp1986} measured the mean positron lifetime over a comparable range; both show that the sigmoidal rise from thermal vacancy formation begins near ${\sim}550\,^\circ$C, with principal sensitivity in the 630--800~$^\circ$C range. At the measured $352\,^\circ$C (625~K), the equilibrium vacancy fraction is $C_\text{eq} \approx 3$--$5 \times 10^{-5}$~ppm ($H_f = 1.31 \pm 0.05$~eV\cite{fluss1980,hehenkamp1986}; $H_f = 1.28$~eV\cite{simmons1963}), so the observed 130--530~ppm population exceeds equilibrium by at least six orders of magnitude. Matching this concentration by Joule heating alone would require raising the specimen above its melting point (1358~K), which is inconsistent with the thermometry, the survival of the foil to 52~A, and the agreement between the LWIR camera and the thermocouple to within 6~$^\circ$C. The rapid, complete recovery of the $S$-parameter to the single-crystal baseline upon current removal further distinguishes the present observations from classical electromigration, which produces permanent mass transport and voiding.

The steady-state balance between current-driven production and thermally activated loss (rather than a frozen equilibrium) fixes the net vacancy production rate at $\Delta C_V/\Delta t \approx 0.6$~ppm\,s$^{-1}$ ($4 \times 10^{16}$~cm$^{-3}$\,s$^{-1}$); combined with the $3.7$~W electrical power input, this bounds the total Frenkel-pair flux to $10^{16}$--$10^{19}$~cm$^{-3}$\,s$^{-1}$.

The sub-minute recombination ($\tau_{\mathrm{rec}} = 1.49 \pm 0.37$~min) is most naturally interpreted as first-order vacancy migration to sinks rather than vacancy--interstitial recombination: Cu interstitials ($E_m = 0.084$~eV\cite{ehrhart1991}) are orders of magnitude more mobile than vacancies ($E_m = 0.70$~eV\cite{ehrhart1991}) and annihilate within nanoseconds of current removal, leaving the PALS-resolved cluster and void population to decay by thermally activated migration to grain boundaries and surfaces as the foil cools.

These observations provide direct experimental evidence that the flash state in a metal is not accounted for by macroscopic Joule heating alone but involves a non-equilibrium, current-induced vacancy population at its core. The ability to electrically sustain vacancy concentrations many orders of magnitude above thermal equilibrium, at temperatures well below the thermal-vacancy onset, introduces the applied current as a thermodynamic control variable for defect populations, with implications for solid-state processing, phase-transformation pathways, and defect engineering of functional solids.

\section*{Methods}

\subsection*{Samples}

Electrolytic copper foils (ASTM B152 C11000, $>99.99$\% purity, $50~\si{\micro\metre}$ thick) were laser-slit using an xTool F1 to $6~\mathrm{mm} \times 50~\mathrm{mm}$ ($0.30~\mathrm{mm}^2$ cross-section), cleaned in dilute HCl, rinsed, and dried. A Cu(100) single crystal etched in nitric acid served as the defect-free reference. Samples were placed into a custom-made four-point probe fixture inside a 6\," Ideal Vacuum cube with a 100~mm CF flange to mate the cube to the beamline. Adjustable copper rods connected the probe to UHV feedthroughs and a power supply; a Picoscope 4260 with a differential amplifier board provided ppm-level recording of current and voltage. Specialized software synchronized the power supply to the experiment.

\subsection*{DBS (WSU 70~keV positron DBS beam line)}

DBS was performed on the WSU variable-energy positron beam (0--73~keV) using a high-purity germanium (HpGe) detector. The $S$- and $W$-parameters were extracted from the 511~keV annihilation photopeak using the standard window definitions of the WSU beamline software: $S$ measures the fraction of counts in the central low-momentum region and $W$ the fraction in the high-momentum wings. The peak count rate in the 511~keV line was ${\sim}700$~counts/s (CuTest3, 3~min integration) to ${\sim}1500$~counts/s (CuTest4, 4~min integration), yielding ${\sim}1.5 \times 10^5$ (CuTest3) and ${\sim}4.1 \times 10^5$ (CuTest4) peak counts per measurement. Positrons at 50~keV are implanted at a mean depth $\bar{z} = (A/\rho)\,E^{1.6} = 2.33$~\si{\micro\metre} in copper ($\rho = 8.96$~g/cm$^3$, $A = 4.0$~\si{\micro\gram/\centi\metre\squared}/keV$^{1.6}$), ensuring bulk sensitivity (Fig.~\ref{fig:technique}c).

CuTest4: 67 measurements over 367~min with current stepped through 23 transitions (0--38~A; ${\sim}5.5$~min cadence including dead time). CuTest5: 38~A~$\to$~0~A decay, 12 points at ${\sim}2.5$~min cadence. CuTest3: 46-point baseline-stability run, no current. Depth scans (0--70~keV): as-received foil (56 points, CuTest6/7; the separate 55-point CuTest2 scan at lower energy resolution is not plotted), post-flash foil (49 points, CuMidl7), and Cu(100) single crystal (132 points, Cu09195).

\subsection*{Resistivity}

Four-point resistivity was logged continuously at 100~Hz during a 0--53~A ramp (140~s) using a GFL data logger with the voltage and current channels calibrated against the power supply readout. Resistivity $\rho$ was computed from the offset-corrected channel voltages using $\rho = (V/I)\times(A/l)$ and converted to $\mu\Omega\cdot\mathrm{cm}$ ($l=30~\mathrm{mm}$ between voltage probes; $A=0.30~\mathrm{mm}^2$).

The CRC thermal baseline in Fig.~\ref{fig:threshold}a was computed from a steady-state 1D energy balance at each current level. The temperature-dependent resistivity of copper follows Matula\cite{matula1979}:
\begin{equation}
\rho(T) = \rho_{293}\bigl[1 + \alpha\,(T - 293)\bigr], \qquad \rho_{293} = 1.678~\si{\micro\ohm.cm},\quad \alpha = 3.93 \times 10^{-3}~\mathrm{K}^{-1}.
\label{eq:rho_cu}
\end{equation}
At each current $I$, the foil temperature $T$ was found by solving
\begin{equation}
I^{2}\,\frac{\rho(T)\,L_\text{gauge}}{A_\text{xs}} = \varepsilon\,\sigma_\text{SB}\,A_\text{surf}\bigl(T^{4} - T_\text{amb}^{4}\bigr) + k_\text{cond}\bigl(T - T_\text{amb}\bigr),
\label{eq:crc_balance}
\end{equation}
where $A_\text{xs} = 0.30~\mathrm{mm}^2$ is the foil cross-section, $L_\text{gauge} = 30~\mathrm{mm}$, $A_\text{surf} = 2(w + t)\,L_\text{tot} = 6.05~\mathrm{cm}^2$ is the radiating surface area ($w = 6~\mathrm{mm}$, $t = 50~\si{\micro\metre}$, $L_\text{tot} = 50~\mathrm{mm}$), $\sigma_\text{SB}$ is the Stefan--Boltzmann constant, $T_\text{amb} = 293$~K, and $k_\text{cond} = 2\kappa_\text{Cu}\,A_\text{xs}/\ell_\text{clip} = 0.024$~W/K accounts for axial conduction to the alligator-clip heat sinks ($\kappa_\text{Cu} = 400$~W\,m$^{-1}$\,K$^{-1}$, clip half-length $\ell_\text{clip} = 10$~mm). The model was evaluated for two emissivity brackets ($\varepsilon = 0.05$ and $0.10$); the predicted resistivity $\rho(T)$ at each $I$ yields the dashed CRC curves in Fig.~\ref{fig:threshold}a.

\subsection*{Temperature measurement}

A follow-up experiment performed on the same sample geometry at the same current levels used three independent techniques to constrain the foil temperature under applied current. All three measurements were taken during sustained flash-state operation at 38~A (after $>12$~min of steady current) in a fixture nominally matching the one used at WSU and NC State for the positron measurements. The three techniques agreed within a ${\sim}30\,^\circ$C band: thermocouple $352 \pm 2\,^\circ$C, LWIR $358 \pm 5\,^\circ$C (after in-situ emissivity calibration against the thermocouple), two-colour pyrometer $382 \pm 15\,^\circ$C; the thermocouple and LWIR camera agree to within $6\,^\circ$C. The per-technique uncertainties combine the manufacturer-specified accuracy with a short-term repeatability estimate from the 38~A hold period.

\textit{Contact thermocouple.} A Type~K thermocouple (Chromel--Alumel, 36~AWG, $127~\si{\micro\metre}$ diameter leads) was capacitance-discharge spot-welded directly to the copper foil at the centre of the gauge section using a bead junction (both wires fused into a single ball prior to welding) to ensure both metals contact the foil at the same equipotential. The TC leads were oriented perpendicular to the current flow direction to reject ohmic (IR-drop) pickup, which would otherwise corrupt the Seebeck voltage: at 38~A the foil voltage gradient is ${\sim}3$~mV/mm, comparable to the ${\sim}12$~mV Seebeck signal at $300\,^\circ$C. As an independent check, the current was briefly reversed during selected measurement intervals; the IR-drop component changes sign under reversal while the Seebeck EMF does not, allowing the two contributions to be separated. Readings taken during the 0~A interludes in the current-cycling protocol (where the IR-drop artifact vanishes) provide additional verification. Temperature was logged with a Digilent USB-2001-TC. The thermocouple recorded $352\,^\circ$C at 38~A after ${>}12$~min of steady current.

\textit{LWIR imaging.} An Optris PI 640i long-wave-infrared camera ($8$--$14~\si{\micro\metre}$) imaged the foil through a 5~mm-thick ZnSe window (Edmund Optics; $>70\%$ transmittance across the LWIR band) fitted to the vacuum cube used for the temperature-confirmation experiment. The vacuum cube was held under rough vacuum during the measurement to minimize convective heat loss from the foil; the LWIR window replaced the CF flange used for the beamline coupling in the positron experiment so that the same foil and clamp geometry could be imaged. Because copper's emissivity is low and strongly oxidation-dependent ($\varepsilon \approx 0.02$--$0.10$), the LWIR camera was not used as a stand-alone thermometer, but rather to (i)~verify that no localized hot spot exceeds the thermocouple reading and (ii)~confirm spatial temperature uniformity along the gauge length. The emissivity setting of the LWIR camera was calibrated against the thermocouple reading at several steady-state current levels to produce a self-consistent thermal map, which recorded $358\,^\circ$C at 38~A.

\textit{Two-colour pyrometry.} A two-colour pyrometer was aimed at the same gauge-section location as the LWIR camera. The two-colour ratio removes the emissivity dependence to first order, at the cost of larger per-measurement noise. The pyrometer read $382\,^\circ$C at 38~A; the 30\,$^\circ$C spread between the three techniques is a realistic estimate of the systematic temperature uncertainty in this geometry.

\subsection*{PALS (NC State PULSTAR positron facility)}

A vacuum bulk PAS system with two re-entrant PMT flanges and a $^{22}$Na source on single-crystal silicon was calibrated against a silicon reference (Extended Data Fig.~\ref{fig:pals_setup}). The timing resolution was ${\sim}300$~ps FWHM (discriminator windows loosened to obtain reasonable count rates). The copper foil (${\sim}2$~mm from source) contributed approximately half the annihilation signal; silicon served as an unchanging internal reference. Each spectrum accumulated $0.66$--$1.98 \times 10^6$ total counts over 7--21~hr per condition. Spectra at four conditions (baseline, 38~A, 13~A, post-flash) were each fitted with three lifetime components using LT10\cite{dryzek1996}, which handles background subtraction internally; reduced $\chi^2$ values ranged from 0.80 to 1.02 (Table~\ref{tab:pals}). For the direct visual comparison of raw spectra (Extended Data Fig.~\ref{fig:pals_spectra}), background noise was subtracted and spectra were peak-normalized. $S$, $W$, and $R$ parameters were independently extracted from the NC State energy spectra and show trends consistent with the WSU DBS results (Fig.~\ref{fig:pals}a--c). Note that the long-lifetime component ($\tau_3$) cannot be straightforwardly converted to void size via the Tao--Eldrup model because positronium is rapidly quenched at metallic surfaces; $\tau_3$ is instead attributed to a surface-state contribution whose intensity tracks the internal surface area created by open-volume defects.

\subsection*{Data analysis}

Per-point $\delta S = 0.002$--0.004 (CuTest4) and 0.005--0.007 (CuTest3) from counting statistics. Trap concentrations: two-state model (Eq.~\ref{eq:concentration}) with specific trapping rate $\mu = 1.1 \times 10^{14}$~s$^{-1}$ (ref.~\cite{mantl1978}), bulk positron lifetime $\tau_b = 110$~ps (i.e.\ $\lambda_b = 9.09 \times 10^{9}$~s$^{-1}$), and DBS contrast $(S_d - S_b)/S_b = 0.07$--$0.10$. With $S_b = 0.396$ and the measured flash-state value $S \approx 0.420$, the two-state model requires $S_d \geq S$, imposing a lower bound $(S_d - S_b)/S_b \geq \Delta S/S_b = 0.024/0.396 = 0.061$; contrasts below this threshold give an unphysical trapping fraction ($\eta > 1$ or $\eta < 0$). This lower bound is larger than the ${\approx}0.04$ value reported for monovacancies\cite{hautojarvi1995,mantl1978,dlubek1979}, indicating a population dominated by vacancy clusters and voids; the upper bound $0.10$ reflects the range of values reported for vacancy clusters and voids in copper. We emphasise that $S_d$ is defined as the saturation value when all positrons trap at the defect, and the values used here are taken from reported cluster/void contrasts rather than measured as a saturation plateau in our own data; if the true $S_d$ for the defect population in our foil lies above this range (i.e., closer to the actual saturation limit), the inferred $C_v$ would be proportionally lower. We therefore quote the 130--530~ppm range as an upper bound. The total dynamic range of vacancy-fraction detection by positron DBS spans roughly ${\sim}10^{-1}$ to ${\sim}10^{3}$~ppm across the progression from sparse monovacancies (detection-limited below ${\sim}0.1$~ppm) to cluster/void saturation (trap-limited above ${\sim}1000$~ppm); the present observation of $>10^{6}\times$ the thermal-equilibrium value at $352\,^\circ$C places the measurement near the upper end of this band, and a positron-background reader should interpret the $10^{6}$ factor with that ceiling in mind. Equilibrium vacancies: $C_\text{eq} = \exp(-H_f / k_B T)$, $H_f = 1.28$~eV (ref.~\cite{simmons1963}).

\textit{Nucleation kinetics.} Two functional forms were used to characterise the 13$\to$38~A transient. (i) An exponential approach to steady state, $S(t) = S_\text{ss} + (S_0 - S_\text{ss})\exp(-t/\tau_\text{nuc})$, gives the time constant $\tau_\text{nuc}$ quoted in the main text (12.1~min at 37~A, 6.7~min at 38~A). (ii) A Boltzmann step, $S(t) = S_\text{low} + \Delta S/(1+\exp(-(t-t_0)/w))$, gives the width $w$ shown in Fig.~\ref{fig:kinetics}a ($w = 2.42 \pm 0.79$~min for the 13$\to$38~A transient). $\tau_\text{nuc}$ and $w$ are different functional-form parameters characterising the same transient and are not related by a simple multiplicative constant; we report both so that readers can reproduce the analysis under either fit form. The rate-scaling exponent is fixed by the two-point data as $\alpha = \ln(\tau_1/\tau_2)/(\Delta J/J_c) = \ln(12.1/6.7)/(1.15-1.12) = 19.7 \approx 20$. Propagating $\pm10\%$ counting-statistics uncertainty on each $\tau_\text{nuc}$ value and a $\pm9\%$ uncertainty on $\Delta J/J_c$ (dominated by the $\pm3$~A uncertainty on $I_c$) gives a total $\delta\alpha/\alpha \approx \pm 30\%$. For extrapolation to $J/J_c = 1.5$, this $\pm 30\%$ on $\alpha$ propagates through $\tau_\text{nuc}(J/J_c) = \tau_\text{ref}\exp[-\alpha(J/J_c - J_\text{ref}/J_c)]$ to roughly a factor-of-8 spread in $\tau_\text{nuc}$ in each direction (plausible range ${\sim}50$~ms to ${\sim}3$~s at $J/J_c = 1.5$), widening at $J/J_c = 2$.

\section*{Data availability}

The complete DBS time series, decay transients, depth scans, PALS spectra, and resistivity logs generated in this study are available as supplementary data accompanying the arXiv preprint.

\section*{Code availability}

Analysis code is available from the corresponding author upon request.

\bmhead{Acknowledgements}

M.H.W.\ led DBS experiments at WSU. M.L.\ designed and operated the PALS system at the NC State PULSTAR positron facility. We thank R.\ Raj for foundational discussions on flash sintering. This research was supported by the Center for Bits and Atoms, MIT.

\section*{Declarations}

\subsection*{Author contributions}

R.F.\ conceived the experiment, led analysis, and wrote the manuscript. L.L.\ and R.N.\ designed sample holders and electrical systems. M.H.W.\ performed all DBS measurements at WSU. M.L.\ performed all PALS and DBS measurements at NC State. N.G.\ supervised the program. All authors discussed results and commented on the manuscript.

\subsection*{Competing interests}

R.F., L.L., R.N., H.A., U.B., and A.C.B.\ are employees and shareholders of General Flash Corp. M.H.W., M.L., and N.G.\ declare no competing interests.

\clearpage
\begin{appendices}
\section*{Extended Data}
\setcounter{figure}{0}
\setcounter{table}{0}
\renewcommand{\thefigure}{\arabic{figure}}
\renewcommand{\thetable}{\arabic{table}}
\renewcommand{\figurename}{Extended Data Fig.}
\renewcommand{\tablename}{Extended Data Table}

\begin{table}[!htbp]
\centering
\caption{\textbf{Summary of experimental runs.} All specimens were electrolytic Cu foils ($0.30~\mathrm{mm}^2$ cross-section) unless noted otherwise. Run identifiers are the lab identifiers used throughout the text. WSU = Washington State University 70\,keV positron beam line; NCSU = NC State PULSTAR positron facility. ``Points'' = number of DBS or resistivity measurements; ``--'' = continuous log.}
\label{tab:runs}
\begin{tabular}{@{}lllllp{0.30\textwidth}@{}}
\toprule
\textbf{Run} & \textbf{Facility} & \textbf{Current} & \textbf{Duration} & \textbf{Points} & \textbf{Purpose} \\
\midrule
CuTest2 & WSU & 0~A & 10.3~hr & 55 & Depth scan of as-received foil \\
CuTest3 & WSU & 0~A & 3.1~hr & 46 & Baseline stability run (50~keV) \\
CuTest4 & WSU & 0--38~A & 6.1~hr & 67 & Current-cycling series (50~keV) \\
CuTest5 & WSU & 38$\to$0~A & 33~min & 12 & Decay transient (50~keV) \\
CuTest6/7 & WSU & 0~A & 10.9~hr & 56 & As-received foil, full 0--70~keV depth scan \\
CuMidl7 & WSU & 0~A & 9.5~hr & 49 & Depth scan of post-flash foil \\
Cu09195 & WSU & 0~A & 25.4~hr & 132 & Cu(100) single-crystal reference depth scan \\
\midrule
Resistivity & WSU & 0--53~A & 140~s & -- & Four-point resistivity ramp \\
PALS & NCSU & 0, 13, 38~A & ${\sim}20$~hr & -- & Lifetime spectra at four conditions (baseline, 38~A, 13~A, post-flash) \\
\bottomrule
\end{tabular}
\end{table}

\begin{figure}[!htbp]
\centering
\includegraphics[width=0.75\textwidth]{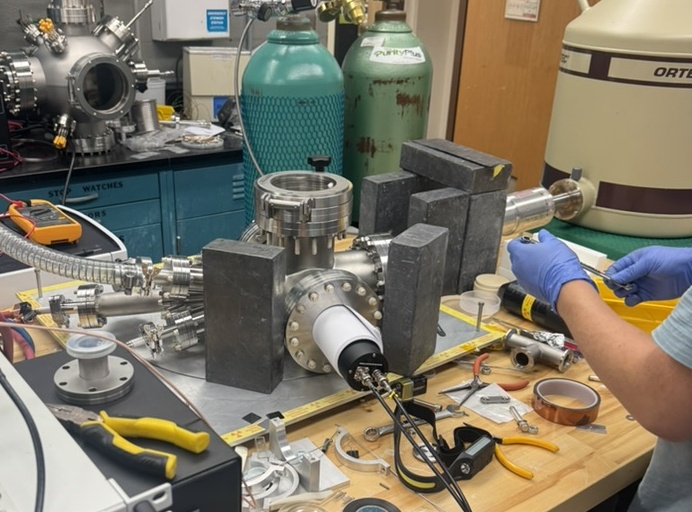}
\caption{\textbf{PALS experimental setup (NC State PULSTAR positron facility), additional view.} Vacuum chamber with lead shielding blocks and ORTEC detectors at the NC State PULSTAR positron facility, complementing the more compact view in Fig.~\ref{fig:technique}f. The $^{22}$Na source on single-crystal silicon is positioned between two re-entrant PMT flanges; the copper foil is held ${\sim}2$~mm from the source and connected to the current supply via a 50~A feedthrough.}
\label{fig:pals_setup}
\end{figure}

\begin{figure}[!htbp]
\centering
\includegraphics[width=0.9\textwidth]{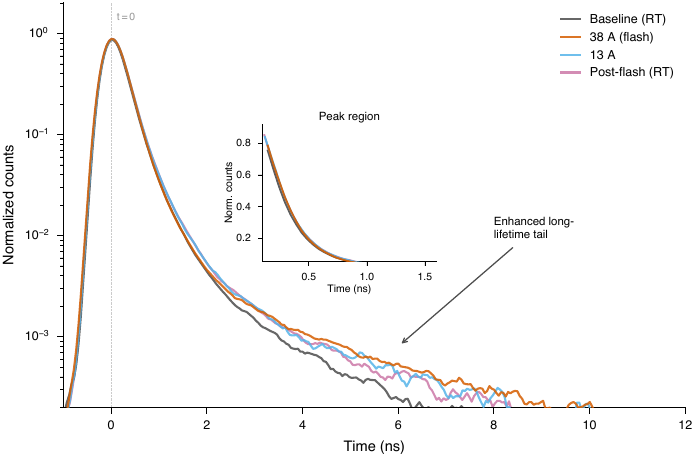}
\caption{\textbf{Raw PALS spectra at each applied-current condition.} Lifetime spectra for the baseline (RT, room-temperature no-current reference), 38~A (flash), 13~A, and post-flash (RT) conditions, background-subtracted and peak-normalized for visual comparison (the lifetime decomposition used in Table~\ref{tab:pals} was instead performed with LT10\cite{dryzek1996}, which handles background subtraction internally). The long-lifetime tail is visibly enhanced at 38~A. Inset: peak region showing that the short components at the baseline and 38~A conditions are similar, while the 13~A and post-flash spectra show slightly longer apparent lifetimes consistent with defect relaxation.}
\label{fig:pals_spectra}
\end{figure}

\end{appendices}


\begin{thebibliography}{10}
\expandafter\ifx\csname url\endcsname\relax
  \def\url#1{\burl{#1}}\fi
\expandafter\ifx\csname urlprefix\endcsname\relax\def\urlprefix{URL }\fi
\providecommand{\bibinfo}[2]{#2}
\providecommand{\eprint}[2][]{\url{#2}}
\providecommand{\doi}[1]{\url{https://doi.org/#1}}
\bibcommenthead

\bibitem{cologna2010}
\bibinfo{author}{Cologna, M.}, \bibinfo{author}{Rashkova, B.} \&
  \bibinfo{author}{Raj, R.}
\newblock \bibinfo{title}{Flash sintering of nanograin zirconia in {$<$5 s} at
  850\,{$^\circ$C}}.
\newblock \emph{\bibinfo{journal}{J. Am. Ceram. Soc.}}
  \textbf{\bibinfo{volume}{93}}, \bibinfo{pages}{3556--3559}
  (\bibinfo{year}{2010}).

\bibitem{biesuz2019}
\bibinfo{author}{Biesuz, M.} \& \bibinfo{author}{Sglavo, V.~M.}
\newblock \bibinfo{title}{Flash sintering of ceramics}.
\newblock \emph{\bibinfo{journal}{J. Eur. Ceram. Soc.}}
  \textbf{\bibinfo{volume}{39}}, \bibinfo{pages}{115--143}
  (\bibinfo{year}{2019}).

\bibitem{raj2012}
\bibinfo{author}{Raj, R.}
\newblock \bibinfo{title}{Joule heating during flash-sintering}.
\newblock \emph{\bibinfo{journal}{J. Eur. Ceram. Soc.}}
  \textbf{\bibinfo{volume}{32}}, \bibinfo{pages}{2293--2301}
  (\bibinfo{year}{2012}).

\bibitem{bamidele2024}
\bibinfo{author}{Bamidele, E.~A.}, \bibinfo{author}{Jalali, S. I.~A.},
  \bibinfo{author}{Weimer, A.~W.} \& \bibinfo{author}{Raj, R.}
\newblock \bibinfo{title}{Flash sintering of tungsten at room temperature
  (without a furnace) in {$<$1 min} by injection of electrical currents at
  different rates}.
\newblock \emph{\bibinfo{journal}{J. Am. Ceram. Soc.}}
  \textbf{\bibinfo{volume}{107}}, \bibinfo{pages}{817--829}
  (\bibinfo{year}{2024}).

\bibitem{k.jha2016beyond}
\bibinfo{author}{JHA, S.~K.}, \bibinfo{author}{TERAUDS, K.},
  \bibinfo{author}{LEBRUN, J.-M.} \& \bibinfo{author}{RAJ, R.}
\newblock \bibinfo{title}{Beyond flash sintering in 3 mol\% yttria stabilized
  zirconia}.
\newblock \emph{\bibinfo{journal}{Journal of the Ceramic Society of Japan}}
  \textbf{\bibinfo{volume}{124}}, \bibinfo{pages}{283–288}
  (\bibinfo{year}{2016}).
\newblock \urlprefix\url{http://dx.doi.org/10.2109/jcersj2.15248}.

\bibitem{das2024reactive}
\bibinfo{author}{Das, S.}, \bibinfo{author}{Durygin, A.},
  \bibinfo{author}{Drozd, V.}, \bibinfo{author}{Sozal, M. S.~I.} \&
  \bibinfo{author}{Cheng, Z.}
\newblock \bibinfo{title}{Reactive flash sintering of {TiZrN} and {TiAlN}
  ternary metal nitrides}.
\newblock \emph{\bibinfo{journal}{Journal of the European Ceramic Society}}
  \textbf{\bibinfo{volume}{44}}, \bibinfo{pages}{2037–2051}
  (\bibinfo{year}{2024}).
\newblock \urlprefix\url{http://dx.doi.org/10.1016/j.jeurceramsoc.2023.11.079}.

\bibitem{raj2016analysis}
\bibinfo{author}{Raj, R.}
\newblock \bibinfo{title}{Analysis of the power density at the onset of flash
  sintering}.
\newblock \emph{\bibinfo{journal}{Journal of the American Ceramic Society}}
  \textbf{\bibinfo{volume}{99}}, \bibinfo{pages}{3226–3232}
  (\bibinfo{year}{2016}).
\newblock \urlprefix\url{http://dx.doi.org/10.1111/jace.14178}.

\bibitem{raj2011fields}
\bibinfo{author}{Raj, R.}, \bibinfo{author}{Cologna, M.} \&
  \bibinfo{author}{Francis, J. S.~C.}
\newblock \bibinfo{title}{Influence of externally imposed and internally
  generated electrical fields on grain growth, diffusional creep, sintering and
  related phenomena in ceramics}.
\newblock \emph{\bibinfo{journal}{J. Am. Ceram. Soc.}}
  \textbf{\bibinfo{volume}{94}}, \bibinfo{pages}{1941--1965}
  (\bibinfo{year}{2011}).

\bibitem{francis2011forging}
\bibinfo{author}{Francis, J. S.~C.} \& \bibinfo{author}{Raj, R.}
\newblock \bibinfo{title}{Flash-sinterforging of nanograin zirconia: field
  assisted sintering and superplasticity}.
\newblock \emph{\bibinfo{journal}{J. Am. Ceram. Soc.}}
  \textbf{\bibinfo{volume}{95}}, \bibinfo{pages}{138--146}
  (\bibinfo{year}{2012}).

\bibitem{narayan2013}
\bibinfo{author}{Narayan, J.}
\newblock \bibinfo{title}{A new mechanism for field-assisted processing and
  flash sintering of materials}.
\newblock \emph{\bibinfo{journal}{Scr. Mater.}} \textbf{\bibinfo{volume}{69}},
  \bibinfo{pages}{107--111} (\bibinfo{year}{2013}).

\bibitem{jo2024}
\bibinfo{author}{Jo, S.}, \bibinfo{author}{Kindelmann, M.},
  \bibinfo{author}{Jennings, D.} \& \bibinfo{author}{Raj, R.}
\newblock \bibinfo{title}{Flash-induced defects in single-crystal {8YSZ}
  characterized by {TEM}, {XRD}, and {R}aman spectroscopy}.
\newblock \emph{\bibinfo{journal}{J. Am. Ceram. Soc.}}
  \textbf{\bibinfo{volume}{107}}, \bibinfo{pages}{5786--5800}
  (\bibinfo{year}{2024}).

\bibitem{jongmanns2018}
\bibinfo{author}{Jongmanns, M.}, \bibinfo{author}{Raj, R.} \&
  \bibinfo{author}{Wolf, D.~E.}
\newblock \bibinfo{title}{Generation of {Frenkel} defects above the {Debye}
  temperature by proliferation of phonons near the {Brillouin} zone edge}.
\newblock \emph{\bibinfo{journal}{New J. Phys.}} \textbf{\bibinfo{volume}{20}},
  \bibinfo{pages}{093013} (\bibinfo{year}{2018}).

\bibitem{jongmanns2020}
\bibinfo{author}{Jongmanns, M.} \& \bibinfo{author}{Wolf, D.~E.}
\newblock \bibinfo{title}{Element-specific displacements in defect-enriched
  {TiO$_2$}: indication of a flash sintering mechanism}.
\newblock \emph{\bibinfo{journal}{J. Am. Ceram. Soc.}}
  \textbf{\bibinfo{volume}{103}}, \bibinfo{pages}{589--596}
  (\bibinfo{year}{2020}).

\bibitem{todd2015}
\bibinfo{author}{Todd, R.~I.} \emph{et~al.}
\newblock \bibinfo{title}{Electrical characteristics of flash sintering:
  thermal runaway of {Joule} heating}.
\newblock \emph{\bibinfo{journal}{J. Eur. Ceram. Soc.}}
  \textbf{\bibinfo{volume}{35}}, \bibinfo{pages}{1865--1877}
  (\bibinfo{year}{2015}).

\bibitem{zhang2017}
\bibinfo{author}{Zhang, Y.}, \bibinfo{author}{Nie, J.}, \bibinfo{author}{Chan,
  J.~M.} \& \bibinfo{author}{Luo, J.}
\newblock \bibinfo{title}{Probing the densification mechanisms during flash
  sintering of {ZnO}}.
\newblock \emph{\bibinfo{journal}{Acta Mater.}} \textbf{\bibinfo{volume}{125}},
  \bibinfo{pages}{465--475} (\bibinfo{year}{2017}).

\bibitem{das2025}
\bibinfo{author}{Das, S.}, \bibinfo{author}{Vukkum, V.~B.},
  \bibinfo{author}{Devaraj, A.}, \bibinfo{author}{Bamidele, E.~A.} \&
  \bibinfo{author}{Raj, R.}
\newblock \bibinfo{title}{Loss of cohesion in metals below the melting point in
  flash-general experiments}.
\newblock \emph{\bibinfo{journal}{J. Am. Ceram. Soc.}}
  \textbf{\bibinfo{volume}{109}} (\bibinfo{year}{2025}).

\bibitem{krause1999}
\bibinfo{author}{Krause-Rehberg, R.} \& \bibinfo{author}{Leipner, H.~S.}
\newblock \emph{\bibinfo{title}{Positron Annihilation in Semiconductors: Defect
  Studies}}  (\bibinfo{publisher}{Springer}, \bibinfo{year}{1999}).

\bibitem{tuomisto2013}
\bibinfo{author}{Tuomisto, F.} \& \bibinfo{author}{Makkonen, I.}
\newblock \bibinfo{title}{Defect identification in semiconductors with positron
  annihilation: Experiment and theory}.
\newblock \emph{\bibinfo{journal}{Rev. Mod. Phys.}}
  \textbf{\bibinfo{volume}{85}}, \bibinfo{pages}{1583--1631}
  (\bibinfo{year}{2013}).

\bibitem{hautojarvi1995}
\bibinfo{author}{Hautoj\"arvi, P.} \& \bibinfo{author}{Corbel, C.}
\newblock \bibinfo{title}{ in \textit{Positron spectroscopy of defects in
  metals and semiconductors}} (eds \bibinfo{editor}{Dupasquier, A.} \&
  \bibinfo{editor}{Mills, A.~P.}) \emph{\bibinfo{booktitle}{Positron
  Spectroscopy of Solids}} \bibinfo{pages}{491--532} (\bibinfo{publisher}{IOS
  Press}, \bibinfo{year}{1995}).

\bibitem{fluss1980}
\bibinfo{author}{Fluss, M.~J.}, \bibinfo{author}{Smedskjaer, L.~C.},
  \bibinfo{author}{Siegel, R.~W.}, \bibinfo{author}{Legnini, D.~G.} \&
  \bibinfo{author}{Chason, M.~K.}
\newblock \bibinfo{title}{Positron annihilation measurement of the vacancy
  formation enthalpy in copper}.
\newblock \emph{\bibinfo{journal}{J. Phys. F: Met. Phys.}}
  \textbf{\bibinfo{volume}{10}}, \bibinfo{pages}{1763--1774}
  (\bibinfo{year}{1980}).

\bibitem{hehenkamp1986}
\bibinfo{author}{Hehenkamp, T.}, \bibinfo{author}{Kurschat, T.} \&
  \bibinfo{author}{L\"uhr-Tanck, W.}
\newblock \bibinfo{title}{Positron lifetime spectroscopy in copper}.
\newblock \emph{\bibinfo{journal}{J. Phys. F: Met. Phys.}}
  \textbf{\bibinfo{volume}{16}}, \bibinfo{pages}{981--987}
  (\bibinfo{year}{1986}).

\bibitem{simmons1963}
\bibinfo{author}{Simmons, R.~O.} \& \bibinfo{author}{Balluffi, R.~W.}
\newblock \bibinfo{title}{Measurement of equilibrium vacancy concentrations in
  copper}.
\newblock \emph{\bibinfo{journal}{Phys. Rev.}} \textbf{\bibinfo{volume}{129}},
  \bibinfo{pages}{1533--1544} (\bibinfo{year}{1963}).

\bibitem{mantl1978}
\bibinfo{author}{Mantl, S.} \& \bibinfo{author}{Triftsh\"auser, W.}
\newblock \bibinfo{title}{Defect annealing studies on metals by positron
  annihilation and resistivity measurements}.
\newblock \emph{\bibinfo{journal}{Phys. Rev. B}} \textbf{\bibinfo{volume}{17}},
  \bibinfo{pages}{1645--1652} (\bibinfo{year}{1978}).

\bibitem{ehrhart1991}
\bibinfo{author}{Ehrhart, P.}
\newblock \bibinfo{title}{ in \textit{Properties and interactions of atomic
  defects in metals and alloys}} (ed.\bibinfo{editor}{Ullmaier, H.})
  \emph{\bibinfo{booktitle}{Atomic Defects in Metals, Landolt-B\"ornstein New
  Series III/25}} \bibinfo{pages}{88--371} (\bibinfo{publisher}{Springer},
  \bibinfo{address}{Berlin}, \bibinfo{year}{1991}).

\bibitem{matula1979}
\bibinfo{author}{Matula, R.~A.}
\newblock \bibinfo{title}{Electrical resistivity of copper, gold, palladium,
  and silver}.
\newblock \emph{\bibinfo{journal}{J. Phys. Chem. Ref. Data}}
  \textbf{\bibinfo{volume}{8}}, \bibinfo{pages}{1147--1298}
  (\bibinfo{year}{1979}).

\bibitem{crc97}
\bibinfo{editor}{Haynes, W.~M.} (ed.) \emph{\bibinfo{title}{CRC Handbook of
  Chemistry and Physics}} \bibinfo{edition}{97th} edn (\bibinfo{publisher}{CRC
  Press}, \bibinfo{address}{Boca Raton, FL}, \bibinfo{year}{2017}).

\bibitem{fulop2026}
\bibinfo{author}{Fulop, R.} \& \bibinfo{author}{Gershenfeld, N.}
\newblock \bibinfo{title}{Critical activation voltage for phonon-mediated
  field-driven phenomena}.
\newblock \bibinfo{howpublished}{Zenodo Preprint} (\bibinfo{year}{2026}).

\bibitem{dryzek1996}
\bibinfo{author}{Dryzek, J.} \& \bibinfo{author}{Kansy, J.}
\newblock \bibinfo{title}{Comparison of three programs: {Positronfit},
  {Resolution} and {LT} for deconvolution of positron lifetime spectra}.
\newblock \emph{\bibinfo{journal}{Nucl. Instr. and Meth. in Phys. Res. A}}
  \textbf{\bibinfo{volume}{380}}, \bibinfo{pages}{576--581}
  (\bibinfo{year}{1996}).

\bibitem{dlubek1979}
\bibinfo{author}{Dlubek, G.}, \bibinfo{author}{Br\"ummer, O.} \&
  \bibinfo{author}{Meyendorf, N.}
\newblock \bibinfo{title}{Impurity-induced vacancy clustering in cold-worked
  nickel}.
\newblock \emph{\bibinfo{journal}{J. Phys. F: Met. Phys.}}
  \textbf{\bibinfo{volume}{9}}, \bibinfo{pages}{1961--1973}
  (\bibinfo{year}{1979}).

\end{thebibliography}
\end{document}